\documentclass[11pt]{article}
\usepackage{latexsym,amssymb}
\def\1bar{1\hskip -.275cm -}
\def\2bar{2\hskip -.275cm -}
\def\3bar{3\hskip -.275cm -}

\newsavebox{\uuunit}
\sbox{\uuunit}
    {\setlength{\unitlength}{0.825em}
     \begin{picture}(0.6,0.7)
        \thinlines
        \put(0,0){\line(1,0){0.5}}
        \put(0.15,0){\line(0,1){0.7}}
        \put(0.35,0){\line(0,1){0.8}}
       \multiput(0.3,0.8)(-0.04,-0.02){10}{\rule{0.5pt}{0.5pt}}
     \end {picture}}

\makeatletter \@addtoreset{equation}{section} \makeatother

\def\bfone{\relax{\rm 1\kern-.35em 1}}

\def\bfone{\relax{\rm 1\kern-.35em 1}}
 
\def\downnormalfill{$\,\,\vrule depth4pt width0.4pt
\leaders\vrule depth 0pt height0.4pt\hfill\vrule depth4pt width0.4pt\,\,$}
\def\WT#1{\mathop{\vbox{\ialign{##\crcr\noalign{\kern3pt}
      \downnormalfill\crcr\noalign{\kern1.8pt\nointerlineskip}
      $\hfil\displaystyle{#1}\hfil$\crcr}}}\limits}

\begin{document}
\begin{titlepage}
\begin{flushright}
DFTT-2003\\
\end{flushright}
\vskip 1.5cm
\begin{center}
{\LARGE {\bf  Perturbative Quantum Gravity in Analogy with}}

{\LARGE {\bf Fermi Theory of Weak Interactions using}}

{\LARGE {\bf Bosonic Tensor Fields }}
\vfill {\large
 Leonardo Modesto} \\
\vfill {
$^1$ Dipartimento di Fisica Teorica, Universit\'a di Torino, \\
$\&$ INFN -
Sezione di Torino\\
via P. Giuria 1, I-10125 Torino, Italy  }
\end{center}
\vfill
\begin{abstract}
{In this paper we 
work in perturbative Quantum Gravity and we introduce a new effective model for gravity. Expanding the Einstein-Hilbert Lagrangian in graviton field powers we have an infinite number of terms. In this paper we study the possibility of an interpretation of more than three graviton interacting vertices as effective vertices of a most fundamental theory that contain tensor fields . Here we introduce a Lagrangian model named I.T.B. (Intermediate-Tensor-Boson) where four gravitational "pseudo-currents" that contain two gravitons couple to three massive tensorial fields of ranks one, three and five respectively. We show that the exchange of those massive particles reproduces, at low energy, the interacting vertices for four or more gravitons. In a particolar version, the model contains a dimensionless coupling constant ``$g$'' and the mass $M_{\Phi}$ of the intermediate bosons as free parameters. The universal gravitational constant  $G_{N}$ is shown to be proportional to the inverse of mass squared of mediator fields, particularly $G_{N}\sim g^{2}/M_{\Phi}^{2}$. A foresighting choice of the dimensionless coupling constant could lower the energy scale where quantum gravity aspects show up. }
\end{abstract}
\end{titlepage}
\section{Introduction to the Model}
In this paper we consider the perturbative expansion of the Einstein-Hilbert action in graviton ($h_{\mu\nu}$) field powers. After the quadratic order that is the free Graviton Lagrangian, the first  not trivial order is the three gravitons interaction. The subsequent  is the four gravitons interaction and so on. The interaction vertex that we study with particular attention is the local four graviton interaction vertex; we show that we can see this term as a "current-current" interaction in analogy with Fermi Theory of Weak Interactions. In a second step we introduce the I.T.B. (Intermediate Tensor Boson) theory for Quantum Gravity in analogy with the I.V.B. (Intermediate Vector Boson) theory that is the non local extension of the Fermi theory. In this Model we introduce an opportune number   
 of Lagrangian terms describing the coupling between currents, made only of gravitons, and tensorial bosons of rank 1, 3, 5. The bosons are the analog of the $W^{+}$ and $W^{-}$ vectors for  I.V.B. theory. \\
Moreover we show that using only another Lagrangian term we can reproduce the tensor structure of all interactions with $5$, $6$, $\dots$ , $n$ gravitons. Finally we compare the I.T.B model with string theory and we propose a new picture where the two theories are two different phases of a more fundamental theory. This can be a theory of massless Tensor Fields  \cite{SF0}, \cite{SF1}, \cite{HSF}, \cite{M01}, \cite{VasilevEq}, \cite{200}, \cite{300}, \cite{400} or a Yang-Mills theory coupled with Sub-Quark  \cite{TRZ1}, \cite{TRZ2}, \cite{WETT}. In the second case String Theory and the I.T.B model could be  two different hadronic phases of a more microscopic sub-quarks theory.
\section{$(\Gamma \, \Gamma)$-Action Gravitons Expansion}
The $(\Gamma \, \Gamma)$ action for  Gravity is 
\begin{equation}
S=-\frac{1}{16\pi G_{N}}\int{G\sqrt{-g}}\,d^4x \label{eq400}
\end{equation}
where       
\begin{equation}
G=g^{\mu \nu}(\Gamma^{\delta}_{\mu \rho}\Gamma^{\rho}_{\nu \delta}-\Gamma^{\rho}_{\mu \nu}\Gamma^{\sigma}_{\rho \sigma})=g^{\mu \nu}g^{\rho \sigma}g^{\tau \varepsilon}(\Gamma_{\sigma \mu \tau}\Gamma_{\varepsilon \nu \rho}-\Gamma_{\sigma\mu \nu}\Gamma_{\varepsilon\rho \tau})
\label{eq401} 
\end{equation}
We consider now the fluctuation of the $g_{\mu \nu}$ metric around the flat background $\eta_{\mu \nu}$ 
\begin{equation}
g_{\mu\nu}=\eta_{\mu\nu}+h_{\mu\nu} \label{eq402}
\end{equation}
This expansion around the flat background is correct only locally. In fact, for Manifold with non trivial topology, the metric can non be globally expanded in $\eta_{\mu \nu}$ plus a fluctuation. This expansion \ref{eq402} is relevant for the description of nature because, in the portion of the Universe where our solar system is placed, the intensity of the fluctuation is about $|h_{\mu\nu}|\sim 10^{-9}$.\\
If we observe the Lagrangian  \ref{eq400} we see that at any order we have only two derivatives.
This is a consequence of the definition of the Cristoffel symbols, in fact such symbols have only low  indices and so the form in terms of $h_{\mu\nu}$ is exact 
\begin{equation}
\Gamma_{\mu \nu \rho}\equiv\frac{1}{2}(\partial_{\rho}h_{\mu\nu}+\partial_{\nu}h_{\mu\rho}-\partial_{\mu}h_{\nu\rho}) \label{eq403}
\end{equation}
Now using the relation $\sqrt{detA}=exp[\frac{1}{2}Tr(logA)]$ we can expand the determinant and the inverse of the metric $g_{\mu\nu}$ in power of $h_{\mu \nu}$ :
\begin{eqnarray}
 \sqrt{det g}& = &\sqrt{det(\delta+h)} = \mbox{exp}\left[\frac{1}{2}Tr(log(\delta+h))\right]=\prod_{n=1}^{\infty}\left\{\sum_{m=0}^{\infty}\frac{1}{m!}\left[\frac{(-1)^{n-1}}{2n}(h^{n})^{\mu}_{\mu}\right]^{m}\right\} =\nonumber \\
&& \hspace{-0.6cm} =1+\frac{1}{2}h^{\mu}\,_{\mu}+\frac{1}{8}(h^{\mu}\,_{\mu})^{2}
           -\frac{1}{4}(h^{2})^{\mu}\,_{\mu} -\frac{1}{8}(h^{\mu}\,_{\mu})(h^{2})^{\mu}\,_{\mu}+\frac{1}{48}(h^{\mu}\,_{\mu})^{3}+\frac{1}{6}(h^{3})^{\mu}\,_{\mu} \nonumber \\
&& \hspace{-0.6cm} -\frac{1}{384}(h^{\mu}\,_{\mu})^{4}-\frac{1}{32}(h^{\mu}\,_{\mu})^{2}(h^{2})^{\mu}\,_{\mu}+\frac{1}{12}(h^{\mu}\,_{\mu})(h^{3})^{\mu}\,_{\mu} -\frac{1}{8}(h^{4})^{\mu}\,_{\mu}+\frac{1}{32}((h^{2})^{\mu}\,_{\mu})^{2} + \dots \nonumber \\
&&
\label{eq405}
\end{eqnarray}
\begin{eqnarray}
(g^{-1})^{\mu\nu}=[(\eta+h)^{-1}]^{\mu\nu}=\eta^{\mu\nu}-h^{\mu\nu}+(h^2)^{\mu\nu}-(h^3)^{\mu\nu}+(h^4)^{\mu\nu}=\sum_{n=0}^{\infty}(-1)^{n}(h^{n})^{\mu\nu} \label{eq407}
\end{eqnarray}
This second relation can be obtained from the Taylor expansion of  $(1+x)^{\alpha}$.
Using those relations the Lorentz gauge condition $\partial^{\mu}h_{\mu\nu}=0$ and, the mass-shell condition ($h^{\mu}\,_{\mu}=0$, $ \Box h_{\mu \nu}$)  the amplitudes can be written
\begin{eqnarray}
&&\hspace{-2.6cm} L_{(4)}=-\frac{1}{64\pi G_{N}}[2h^{\mu\nu}\partial_{\tau}h_{\nu\rho}h^{\tau\sigma}\partial_{\mu}h_{\sigma}\,^{\rho}-2h^{\mu\nu}\partial_{\tau}h_{\nu\rho}h^{\tau\sigma}\partial_{\sigma}h_{\mu}\,^{\rho}+\nonumber \\
&& +3h^{\mu\nu}\partial_{\tau}h_{\nu\rho}h^{\rho\sigma}\partial_{\mu}h^{\tau}\,_{\sigma}+\frac{3}{2}h^{\mu\nu}\partial_{\tau}h_{\nu\rho}h^{\rho\sigma}\partial^{\tau}h_{\mu\sigma}+\nonumber \\
&& +2h^{\mu\nu}\partial_{\tau}h_{\nu\rho}h_{\mu\sigma}\partial^{\sigma}h^{\tau\rho}+h^{\mu\nu}\partial_{\tau}h_{\mu\rho}h_{\nu\sigma}\partial^{\rho}h^{\tau\sigma}+\nonumber \\
&& +\frac{1}{2}h^{\mu\nu}\partial_{\tau}h_{\mu\rho}h_{\nu\sigma}\partial^{\tau}h^{\rho\sigma}+h^{\mu\nu}\partial_{\mu}h_{\tau\rho}h_{\nu\sigma}\partial^{\rho}h^{\tau\sigma}+\nonumber \\
&& +\frac{3}{2}h^{\mu\nu}\partial_{\mu}h_{\tau\rho}h_{\nu\sigma}\partial^{\sigma}h^{\tau\rho}+\nonumber \\
&& +\frac{5}{2}h^{\mu\nu}\partial^{\tau}h^{\rho\sigma}h_{\mu\sigma}\partial_{\tau}h_{\rho\nu}-\frac{5}{2}h^{\mu\nu}\partial^{\tau}h^{\rho\sigma}h_{\mu\tau}\partial_{\nu}h_{\rho\tau}+\nonumber \\
&& +\frac{13}{2}h^{\mu\nu}\partial^{\tau}h^{\rho\sigma}h_{\mu\nu}\partial_{\tau}h_{\rho\sigma}+\frac{3}{2}h^{\mu\nu}\partial^{\tau}h^{\rho\sigma}h_{\mu\nu}\partial_{\sigma}h_{\rho\tau}+\nonumber \\
&& +\frac{29}{2}h^{\mu\nu} \partial_{\tau} h_{\mu\nu} h^{\rho\sigma}\partial^{\tau}  h_{\rho\sigma}]
 \label{eq436}
\end{eqnarray}
Since we are studying the tree level gravitons interaction it is sufficient to consider only the on shell 
Lagrangian terms. In fact all the interactions that contain the tensors  $\partial_{\mu}h_{\mu\nu}$, $h^{\mu}\,_{\mu}$ and $\Box h_{\mu \nu}$, produce amplitudes proportional to  $k^{\mu}\epsilon_{\mu\nu}$, $\epsilon^{\mu}\,_{\mu}$ and $k^2$ that are identically zero on external gravitons states.\\
Now we introduce tensors with tree and five indexes, that we call "tensor currents" and we rewrite the Lagrangian (\ref{eq436}) as a "current-current" interaction in analogy with Fermi theory of the weak interactions. The currents need to do it are 
\begin{eqnarray}
&J_{(1)}^{\mu}=(\partial^{\mu}h_{\nu\rho})h^{\nu \rho}\\
&J_{(2)}^{\mu\nu\rho}=(\partial^{\sigma}h^{\mu\nu})h_{\sigma}\,^{\rho}\\
&J_{(3)}^{\mu\nu\rho\sigma\delta}=(\partial^{\mu}h^{\nu\rho})h^{\sigma\delta} 
\label{eq413}
\end{eqnarray}
Using those currents in the Lagrangian (\ref{eq436}) we obtain this compact  form for the four gravitons interaction 
\begin{equation}
L_{4}=-\frac{1}{64\pi G_{N}}[\sum_{i,j=1}^{2}a_{ij}J_{i}J_{j}+a_{(3)}J_{(3)}J_{(3)}]               \label{eq414}
\end{equation}

\section{$G_{N}$ Constant Expansion}
The fundamental constant in gravity is the Newton constant $G_{N}$ that in natural units $c = 1$ and $\hbar = 1$ is 
\begin{equation}
G_{N}=\frac{1}{M^{2}_{P}}\sim (10^{19}GeV)^{-2}  
\label{eq415}
\end{equation}
In front of the Lagrangian we have the factor $-\frac{1}{64\pi G_{N}}$, and we can reabsorb it in the definition of the graviton field :
\begin{equation}
h_{\mu\nu}\rightarrow \sqrt{64\pi G_{N}}h_{\mu\nu}
\end{equation}
In this way we obtain mass dimension for the graviton $h_{\mu \nu}$ while the metric is dimensionless. With this definition we obtain a factor  $(64\pi G_{N})^{\frac{n}{2}}$ in front of the $n$-oder interaction in the gravitons expansion of the Lagrangian.
The scheme of the Lagrangian with correct powers of $G_N$ is 
\begin{eqnarray}
&& \hspace{-0.5cm}L=(\partial h\partial h)+\sqrt {G_{N}}(\partial h\partial h)h+(\sqrt{G_{N}})^{2}(\partial h\partial h)hh\nonumber \\
&& +(\sqrt{G_{N}})^{3}(\partial h\partial h)hhh+ (\sqrt{G_{N}})^{4}(\partial h\partial h)hhhh \cdots = \nonumber \\
&& =\sum_{n=0}^{\infty}[(G_{N})^{\frac{n}{2}}(\partial h\partial h)(h)^{n}] 
\end{eqnarray}
Where $64\pi$ factors are understood.\\
At this point we can rewrite the Lagrangian \ref{eq414} in the following way :
\begin{eqnarray}
L_{4}&=-64\pi G_{N}[\sum_{i,j=1}^{2}a_{ij}J_{i}J_{j}+a_{(3)}J_{(3)}J_{(3)}]=\\
      &=-\frac{64\pi}{M_{P}^{2}}[\sum_{i,j=1}^{2}a_{ij}J_{i}J_{j}+a_{(3)}J_{(3)}J_{(3)}]
\label{eq416}
\end{eqnarray}

\section{The I.T.B Theory for Quantum Gravity}
As said in the first section the I.V.B theory replaces the Fermi Theory of the weak interactions. In this new model the fermionic local interaction becomes a non local interaction with the exchange of a massive boson ($W^{+}$ or $W^{-}$). The mass of the bosons is exactly the inverse of the $G_F$ Fermi constant ($(G_F)^{-1/2} = M_W$). Now we introduce for Gravity an analog of the I.V.B model that we call I.T.B (Intermediate Tensor Boson Model). \\
In the I.T.B Model the analog of the $W_{\mu}^{+}$ $W_{\mu}^{-}$ bosons are tensor fields of rang 
$1$, $3$ and $5$ : 
\begin{equation}  
\Phi_{\mu} \hspace{1cm} \Phi_{\mu\nu\rho} \hspace{1.0cm} \Phi_{\mu\nu\rho\sigma\tau}
\end{equation} 
Such fields couple to what we call "gravitational currents", in analogy with the Fermi model, through the following Lagrangian terms with a dimensionless coupling constant, as in the I.V.B model.
\begin{eqnarray}
&& L_{(0)} = gJ_{(0)}^{\mu} \Phi_{\mu} = gh^{\nu \rho} \partial^{\mu}h{\nu \rho} \Phi_{\mu}\nonumber \\
&& L_{(1)}=gJ_{(1)}^{\mu\nu\rho}\,\Phi_{\mu\nu\rho}=g(\partial^{\mu}h^{\nu\sigma})h_{\sigma}\,^{\rho}\Phi_{\mu\nu\rho} \nonumber \\
&& L_{(2)}=gJ_{(2)}^{\mu\nu\rho}\,\Phi_{\mu\nu\rho}=g(\partial^{\sigma}h^{\mu\nu})h_{\sigma}\,^{\rho}\Phi_{\mu\nu\rho}\nonumber \\
&& L_{(3)}=gJ_{(3)}^{\mu\nu\rho\sigma\tau}\,\Phi_{\mu\nu\rho\sigma\tau}=g(\partial^{\mu}h^{\nu\rho})h^{\sigma\tau}\Phi_{\mu\nu\rho\sigma\tau}
\label{eq417}
\end{eqnarray}
Those Lagrangian terms reproduce all the local four graviton interactions as a low energy limit  ($M^{2}_{\Phi}\gg k^{2}_{\Phi}$) of amplitudes in which gravitons interact through the exchange of  massive fields $\Phi_{\mu}$, $\Phi_{\mu\nu\rho}$ or $\Phi_{\mu\nu\rho\sigma\tau}$.\\
We introduce now the following tensor factors for the propagetors of the higher spin fields 
\begin{eqnarray}
&&\hspace{2.2cm}K^{(1)}_{\mu,\nu}=\frac{29}{2}\eta_{\mu\nu}  \hspace{7.8cm}
\label{eq439}
\end{eqnarray}
\begin{eqnarray} 
\label{eq437}
K^{(3)}\,_{\mu\nu\rho,\alpha\beta\gamma}&=&\frac{1}{2}\eta_{\mu\alpha}\eta_{\nu\beta}\eta_{\rho\gamma}+\frac{3}{2}\eta_{\mu\alpha}\eta_{\nu\gamma}\eta_{\rho\beta}
+\eta_{\mu\beta}\eta_{\nu\alpha}\eta_{\rho\gamma}+\nonumber \\
&-&\frac{15}{8}\eta_{\mu\gamma}\eta_{\nu\alpha}\eta_{\rho\beta}    
+\frac{5}{8}\eta_{\mu\beta}\eta_{\nu\gamma}\eta_{\rho\alpha}+\frac{7}{8}\eta_{\mu\gamma}\eta_{\nu\beta}\eta_{\rho\alpha}   
\label{eq418} 
\end{eqnarray}
\begin{eqnarray}
K^{(5)}\,_{\mu_{1}\nu_{1}\rho_{1}\delta_{1}\sigma_{1}\, , \, \mu_{2}\nu_{2}\rho_{2}\delta_{2}\sigma_{2}}&=&\, \frac{13}{2}  \eta_{\mu_{1}\mu_{2}} \eta_{\nu_{1}\nu_{2}} \eta_{\rho_{1}\rho_{2}} \eta_{\delta_{1}\delta_{2}} \eta_{\sigma_{1}\sigma_{2}}+ \nonumber \\
&+&\frac{3}{2} \eta_{\mu_{1}\mu_{2}} \eta_{\nu_{1}\rho_{2}} \eta_{\rho_{1}\nu_{2}} \eta_{\delta_{1}\delta_{2}} \eta_{\sigma_{1}\sigma_{2}}+\nonumber \\
&-&\frac{5}{2} \eta_{\mu_{1}\mu_{2}} \eta_{\nu_{1}\nu_{2}} \eta_{\rho_{1}\sigma_{2}} \eta_{\delta_{1}\delta_{2}} \eta_{\sigma_{1}\rho_{2}}+\nonumber \\
&+&\frac{5}{2} \eta_{\mu_{1}\mu_{2}} \eta_{\nu_{1}\sigma_{2}} \eta_{\rho_{1}\rho_{2}} \eta_{\delta_{1}\delta_{2}} \eta_{\sigma_{1}\nu_{2}}
\label{eq438}
\end{eqnarray}
We start with the field of rank $3$. In this case the Lagrangian is $L_{(1)}+L_{(2)}$ and we  calculate the scattering amplitudes for $4$ gravitons using the $S$-matrix.\\
We recall that the $S$-matrix connects the final state $|\Phi(t=+\infty) \rangle$ to the initial state 
$|\Phi(t=-\infty) \rangle$ defined by 
\begin{equation}
|\Phi(t=+\infty) \rangle=S|\Phi(t=-\infty) \rangle 
\end{equation}
The probability to obtain the system in the general state $|f \rangle$ after the scattering, is 
\begin{equation}
|\langle f |\Phi(t=+\infty) \rangle |^{2} \label{eq419}
\end{equation}
and the unitarity of the $S$-matrix can be written as
\begin{equation}
\sum_{f}|S_{fi}|^{2}=1   \label{eq421}
\end{equation}
The general form of the $S$-matrix that has those property is 
\begin{equation}
S=\sum_{n=0}^{\infty} \frac{(i)^{n}}{n!} \int \dots \int d^{4}x_{1} d^{4}x_{2} \dots  d^{4}x_{n}T[:L_{I}(x_{1})::L_{I}(x_{2}):\dots:L_{I}(x_{n}):]
\end{equation}
Now we consider the second order in the $S$-matrix for the Lagrangian $L_{1}+ L_{2}$, with $L_{1}$ and $L_{2}$ defined in \ref{eq417} :
\begin{eqnarray}
S^{(2)}&=&\frac{(i)^{2}}{2!} \int \int d^{4}x_{1} d^{4}x_{2} T[:(L_{1}+ L_{2})(x_{1})::(L_{1}+ L_{2})(x_{2}):] \nonumber \\
      &=&-\frac{1}{2}\int d^{4}x_{1} d^{4}x_{2}g^{2}[:(J_{(1)}^{\mu\nu\rho}\WT{\Phi_{\mu\nu\rho})(x_{1})::(J_{(1)}^{\sigma\delta\tau}\Phi}\hspace{-0.1cm}\vphantom{.}_{\sigma\delta\tau})(x_{2}): \nonumber \\
&&\hspace{2.5cm} +2:(J_{(1)}^{\mu\nu\rho}\WT{\Phi_{\mu\nu\rho})(x_{1})::(J_{(2)}^{\sigma\delta\tau}\Phi}\hspace{-0.1cm}\phantom{.}_{\sigma\delta\tau})(x_{2}): \nonumber \\
&& \hspace{2.7cm} +:(J_{(2)}^{\mu\nu\rho}\WT{\Phi_{\mu\nu\rho})(x_{1})::(J_{(2)}^{\sigma\delta\tau}\Phi}
\hspace{-0.1cm}\vphantom{.}_{\sigma\delta\tau})(x_{2}):]       
\label{eq422}
\end{eqnarray}
Introducing the propagator for the field $\Phi_{\mu\nu\rho}$ in $S$ at the second order  and taking $M^{2}_{\Phi} \gg k^{2}_{\Phi}$ we obtain the local amplitude for $4$ gravitons.
 This amplitude is exactly what we obtain from Einstein action (\ref{eq436}) at first order in the $S$-matrix. In this section we don't give calculations detail report the calculus but only the idea of the mechanism (see the Appendix).  
 Using Feyman diagrams we can depict the interaction at high ($M^{2}_{\Phi}\sim k^{2}_{\Phi}$) and low ($M^{2}_{\Phi} \gg k^{2}_{\Phi}$) energy, and this is very similar to what we find when we pass from I.V.B model to Fermi theory provided the following identifications 
\begin{eqnarray}
&h_{\mu\nu} \leftrightarrow \psi \\
&\Phi_{\mu\nu\rho} \leftrightarrow W^{+}_{\mu}\, , \, W^{-}_{\mu}\\
&G_{N} \leftrightarrow G_{F}        
\label{eq423}
\end{eqnarray}
Qualitatively the $S$-matrix at the second order is :
\begin{eqnarray}
S &\sim& g^{2} [(\partial h)h\WT{\Phi](x_{1})[(\partial h)h\Phi}](x_{2}] \sim \nonumber \\
   &\sim&  g^{2}(\partial h)h \frac{1}{K^{2}_{\Phi}-M^{2}_{\Phi}}(\partial h)h\rightarrow\nonumber \\
   &\rightarrow& g^{2}(\partial h)h\frac{1}{M^{2}_{\Phi}}(\partial h)h \sim \nonumber \\
   &\sim& G_{N} (\partial h)h(\partial h)h
\label{eq424}
\end{eqnarray}
At this point we analyse the result from the I.T.B model. \\
In this model we can interpret the Newton constant $G_N$ as proportional to the mass of the fields of rank $1$, $3$, and $5$, $\phi_{\mu}$,  $\phi_{\mu\nu\rho}$ and $\phi_{\mu\nu\rho\sigma\tau}$ that are the mediators of the gravitational interaction.  
\begin{equation}
G_{N}=\frac {g^{2}}{M_{\Phi}^{2}}
\end{equation}
The other important fact is that we can have a lower energy scale where gravity takes a quantum nature. In the low energy limit the non-local amplitude becomes local and comparing with the Einstein theory we obtain the following identification 
\begin{equation}
\frac{g^{2}}{M_{\Phi}^{2}} \sim G_{N} \sim (10^{-19})^{2}\,Gev^{-2}
\end{equation}
or :
\begin{equation}
\frac{g}{M_{\Phi}} \sim \sqrt G_{N} \sim 10^{-19}\,Gev^{-1}
\end{equation}
If the adimentional coupling constant "$g$" is very small ($g \ll 1$), we can apply the perturbative theory and for $g \sim 10^{-4}$ we obtain $M_{\Phi} \sim 10^{15}\, GeV$, that is the Grand Unification scale. If $g \sim 10^{-15} $ than $M_{\Phi} \sim 10^{4}\, Gev$ and we cauld see Quantum Gravitational effects at accelerators.

\subsection{Tensor Fields as Reducible Representation of the Poincar\'e Group} 
\label{RedPoin}
The tensor factors \ref{eq439}, \ref{eq418} e \ref{eq438} inside the propagators
have no definite symmetry. Now we rewrite the tensor factors \ref{eq418}, \ref{eq438} as a sum of propagators for other fields, with always rank $3$ and $5$, that are reducible representations of the Poincar\'e group with definite tensor symmetry.  We can write the tensors  $K^{(1)},\,K^{(2)},\,K^{(3)}$ as following.\\
 For $K^{(3)}$ we have 
\begin{eqnarray}
K^{(3)}\,_{\mu\nu\rho,\alpha\beta\gamma}&=&\frac{1}{2}\eta_{\mu\alpha}\eta_{\nu\beta}\eta_{\rho\gamma}+\frac{1}{2}\eta_{\mu\alpha}\eta_{\nu\gamma}\eta_{\rho\beta}+\\
&&+\eta_{\mu\alpha}\eta_{\nu\gamma}\eta_{\rho\beta}-\eta_{\mu\beta}\eta_{\nu\gamma}\eta_{\rho\alpha}+\\
&&+\eta_{\mu\beta}\eta_{\nu\alpha}\eta_{\rho\gamma}-\eta_{\mu\gamma}\eta_{\nu\alpha}\eta_{\rho\beta}+\\
&&+\frac{7}{8}\eta_{\mu\gamma}\eta_{\nu\beta}\eta_{\rho\alpha}-\frac{7}{8}\eta_{\mu\gamma}\eta_{\nu\alpha}\eta_{\rho\beta}+\\
&&+\frac{13}{8}\eta_{\mu\beta}\eta_{\nu\gamma}\eta_{\rho\alpha}
\label{eq440}
\end{eqnarray}
Any line in the tensor (\ref{eq440}) represents the propagator of a field that can be decomposed in irreducible representations of the Poncar\'e group.\\ 
For $K^{(5)}$ we have the following decomposition 
\begin{eqnarray}
K^{(5)}_{\mu_{1}\nu_{1}\rho_{1}\delta_{1}\sigma_{1}\, , \, \mu_{2}\nu_{2}\rho_{2}\delta_{2}\sigma_{2}}&=&\frac{13}{2}(\eta_{\mu_{1}\mu_{2}} \eta_{\nu_{1}\nu_{2}} \eta_{\rho_{1}\rho_{2}} \eta_{\delta_{1}\delta_{2}} \eta_{\sigma_{1}\sigma_{2}}+ \eta_{\mu_{1}\mu_{2}} \eta_{\nu_{1}\rho_{2}} \eta_{\rho_{1}\nu_{2}} \eta_{\delta_{1}\delta_{2}} \eta_{\sigma_{1}\sigma_{2}}) \nonumber \\
&&+\frac{5}{2}(\eta_{\mu_{1}\mu_{2}} \eta_{\nu_{1}\nu_{2}} \eta_{\rho_{1}\rho_{2}} \eta_{\delta_{1}\delta_{2}} \eta_{\sigma_{1}\sigma_{2}}-\eta_{\mu_{1}\mu_{2}} \eta_{\nu_{1}\nu_{2}} \eta_{\rho_{1}\sigma_{2}} \eta_{\delta_{1}\delta_{2}} \eta_{\sigma_{1}\rho_{2}})\nonumber \\
&&+\frac{5}{2}(\eta_{\mu_{1}\mu_{2}} \eta_{\nu_{1}\nu_{2}} \eta_{\rho_{1}\rho_{2}} \eta_{\delta_{1}\delta_{2}} \eta_{\sigma_{1}\sigma_{2}}+\eta_{\mu_{1}\mu_{2}} \eta_{\nu_{1}\sigma_{2}} \eta_{\rho_{1}\rho_{2}} \eta_{\delta_{1}\delta_{2}} \eta_{\sigma_{1}\nu_{2}})
\label{eq442}       
\end{eqnarray}

\section{Interaction with $5$ Gravitons (Tensor Structure)} 
The interaction vertexes that contain $5, 6, \dots n$ gravitons have a tensor structure such that we can obtain those interactions introducing one or plus gravitons into the four graviton vertex.
To carry out this idea we must introduce another Lagrangian term to the I.T.B model; such term couples two massive fields with a graviton. This Interaction term in the low energy limit reproduces the tensor behaviour of the local interactions of the Einstein action.\\
For example we consider a typical $5$-gravitons interaction that we can obtain from the expansion of the Einstein-Hilbert action 
\begin{eqnarray} 
L_{(5)grav} = (32 \pi G_{N})^{\frac{3}{2}} \partial_{\mu}h_{\nu\sigma}h^{\nu}\,_{\tau}h^{\mu}\,_{\delta} \partial^{\delta}h^{\epsilon\sigma}h_{\epsilon}\,^{\tau} = (32 \pi G_{N})^{\frac{3}{2}} J_{\mu\sigma\tau}h^{\mu}\,_{\delta} J^{\delta\sigma\tau}
\label{eq427}
\end{eqnarray}
The Lagrangian term that we must introduce in the I.T.B model to reproduce (\ref{eq427}) is 
\begin{equation}
L_{(4)}=gM_{\Phi}h_{\mu\nu}\Phi^{\mu\rho\sigma}\Phi^{\nu}\, _{\rho\sigma}  \label{eq428}
\end{equation}
Omitting the indexes, the total lagrangian is 
\begin{equation}
L_{T}=L_{(1)}+L_{(4)}=gJ_{(1)}\Phi+gM_{\Phi}h\Phi\Phi  \label{eq429}
\end{equation}
To calculate the $5$ gravitons amplitude in the I.T.B model, we consider the third order in the $S$-matrix at tree level :
\newpage
\begin{eqnarray}
&&\hspace{-0.8cm}S^{(3)}=\frac{(i)^{3}}{3!} \int d^{4}x_{1}d^{4}x_{2}d^{4}x_{3}T[:L(x_{1})::L(x_{2})::L(x_{3}):] \nonumber \\
&&=\frac{(i)^{3}}{3!} \int d^{4}x_{1}d^{4}x_{2}d^{4}x_{3}T[:(gJ^{\mu_{1}\nu_{1}\rho_{1}}\Phi_{\mu_{1}\nu_{1}\rho_{1}}+gM_{\Phi}h_{\mu_{1}\nu_{1}}\Phi^{\mu_{1}\rho_{1}\sigma_{1}}\Phi^{\nu_{1}}\,_{\rho_{1}\sigma_{1}})(x_{1}): \nonumber \\
&&\hspace{4.5cm}:(gJ^{\mu_{2}\nu_{2}\rho_{2}} \Phi_{\mu_{2}\nu_{2}\rho_{2}}+gM_{\Phi}h_{\mu_{2}\nu_{2}}\Phi^{\mu_{2}\rho_{2}\sigma_{2}}\Phi^{\nu_{2}}\,_{\rho_{2}\sigma_{2}})(x_{2}): \nonumber \\
&&\hspace{4.5cm}:(gJ^{\mu_{3}\nu_{3}\rho_{3}} \Phi_{\mu_{3}\nu_{3}\rho_{3}}+gM_{\Phi}h_{\mu_{3}\nu_{3}}\Phi^{\mu_{3}\rho_{3}\sigma_{3}}\Phi^{\nu_{3}}\,_{\rho_{3}\sigma_{3}})(x_{3}):]\nonumber \\
&&=\frac{4(i)^{3}}{3}\int d^{4}x_{1}d^{4}x_{2}d^{4}x_{3}g^{3}M_{\Phi} \times \nonumber \\
&&\hspace {1.5cm} \times :(J_{(1)}^{\mu\nu\rho}\WT{\Phi_{\mu\nu\rho})(x_{(1)})::(h_{\alpha\beta}\Phi}\hspace{-0.1cm}\vphantom{.}^{\alpha\gamma\delta}\WT{\Phi^{\beta}\,_{\gamma\delta})(x_{(2)}):
        :(J_{(1)}^{\epsilon\tau\sigma}\Phi}\hspace{-0.1cm}\vphantom{.}_{\epsilon\tau\sigma})(x_{(1)}): + \dots \nonumber \\
&&=\frac{(4i)^{3}}{3} \int d^{4}x_{1}d^{4}x_{2}d^{4}x_{3}g^{3}M_{\Phi}J_{(1)}^{\mu\nu\rho}(x_{1}) \int \frac{d^{4}k_{1}}{(2\pi)^{4}}(-i)\frac{e^{ik_{1}(x_{2}-x_{1})}}{k_{1}^{2}-M_{\Phi}^{2}}(\delta^{\alpha}_{\mu}\delta^{\gamma}_{\nu}\delta^{\delta}_{\rho}) \times \nonumber \\
&&\hspace{1.5cm} h_{\alpha\beta}(x_{2})\int \frac{d^{4}k_{2}}{(2\pi)^{4}}(-i)\frac{e^{ik_{2}(x_{3}-x_{2})}}{k_{2}^{2}-M_{\Phi}^{2}}(\delta^{\beta}_{\epsilon}\eta^{\gamma}_{\tau}\eta^{\sigma}_{\delta})J_{(1)}^{\epsilon\tau\sigma}(x_{3}) + \dots \nonumber \\
&&=g^{3}M_{\Phi}\frac{4 i}{3}\int d^{4}x_{1}d^{4}x_{2}d^{4}x_{3}
        \int \frac {d^{4}k_{1}}{(2\pi)^{4}} \int \frac {d^{4}k_{1}}{(2\pi)^{4}}
        \frac{e^{ik_{1}(x_{2}-x_{1})}}{k_{1}^{2}-M_{\Phi}^{2}} \frac{e^{ik_{2}(x_{3}-x_{2})}}{k_{2}^{2}-M_{\Phi}^{2}}\times \nonumber \\
&&\hspace{1.5cm}\times J_{(1)}^{\mu\nu\rho}(x_{1})h_{\mu\epsilon}(x_{2})J_{(1)}\,^{\epsilon}\,_{\nu\rho}(x_{3}) + \dots  \rightarrow \nonumber \\
&&\rightarrow \frac{4}{3} \frac{g^{3}}{M_{\phi}^{3}} \int d^{4}x_{1}J_{(1)}^{\mu\nu\rho}h_{\mu}^{\epsilon}J_{(1)\epsilon\nu\rho}(x_{1}) + \dots 
\label{eq430}
\end{eqnarray}
Comparing the result with the Einstein action we obtain $(32\pi G_{N})^{\frac{3}{2}}=\frac{4}{3}\frac{g^{3}}{M_{\Phi}^{3}}$ and so 
$G_{N}\sim \frac{g^{2}}{M_{\Phi}^{2}}$.

\section{"$n$" Graviton Interaction (Tensor Structure)}
In the previous section we have introduced another Lagrangian term that reproduces all the interactions for $5$ or  in general "$n$" gravitons ($n \geq 6$) at least in the tensor structure. Now 
we omit the tensor indexes and we explain the mechanism.
A typical Lagrangian term  for $6$ gravitons that we obtain from the Einstein action is :
\begin{equation}
L_{6grav} \sim (G_{N})^{2} (\partial h)h(\partial h)hhh
\end{equation}
This local interaction can be reproduced with the I.T.B effective Lagrangian 
\begin{equation}
L_{T}=L_{(1)}+L_{(2)}+L_{(3)}+L_{(4)}
\end{equation}
where $L_{(1)}, L_{(2)}, L_{(3)}, L_{(4)}$ were defined before. \\
At fourth order we obtain the $6$ graviton interaction 
\begin{eqnarray}
S^{(4)}&=&\frac{(i)^{2}}{(4!)} \int d^{4}x_{1} d^{4}x_{2} d^{4}x_{3} d^{4}x_{4}T[:L_{T}(x_{1})::L_{T}(x_{2})::L_{T}(x_{3})::L_{T}(x_{4}):] \sim \nonumber \\
&&\hspace{1.0cm}\sim (g \partial h h \WT{ \Phi)(x_{1})(gM_{\Phi}h\Phi} \WT{\Phi)(x_{2})(gM_{\Phi}h\Phi}\WT{\Phi)(x_{3})(g \partial h h \Phi})(x_{4})\sim \nonumber \\
&&\hspace{1.0cm}\sim g^{4}M_{\Phi}^{2} (\partial h)h \frac{1}{K_{1}^{2}-M_{\Phi}^{2}} h \frac{1}{K_{2}^{2}-M_{\Phi}^{2}} h \frac{1}{K_{3}^{2}-M_{\Phi}^{2}} (\partial h)h\,\rightarrow \nonumber \\
&&\hspace{1.0cm}\rightarrow \frac{g^{4}}{M_{\Phi}^{2}}(\partial h)h(\partial h)hhh
\label{eq446}
\end{eqnarray}
For the "$n$" graviton interaction we must take the $S$-matrix at the "$n-2$" order :
\begin{eqnarray}
S^{(n-2)}& \sim &(g \partial hh \WT{\Phi)(x_{1})(gM_{\Phi}h\Phi} \WT{\Phi)(x_{2})(gM_{\Phi}h\Phi}\WT{\Phi)(x_{3})(gM_{\Phi}h\Phi} \Phi)(x_{4})\dots \nonumber \\
&&\dots  (gM_{\Phi}h\Phi\WT{ \Phi)(x_{n-2})(gM_{\Phi}h\Phi}\WT{ \Phi)(x_{n-1})(g \partial hh\Phi})(x_{n})\sim \nonumber \\
&&\sim g^{n}M_{\Phi}^{n-2} \partial hh\frac{1}{K_{1}^{2}-M_{\Phi}^{2}}h\frac{1}{K_{2}^{2}-M_{\Phi}^{2}}h\frac{1}{K_{3}^{2}-M_{\Phi}^{2}}\dots \frac{1}{K_{n-1}^{2}-M_{\Phi}^{2}}\partial hh \rightarrow \nonumber \\
&& \rightarrow g^{n}\frac{M_{\Phi}^{n-2}}{M_{\Phi}^{2(n-1)}}\partial hhhh\dots\dots\partial hh= \nonumber \\
&&=\frac{g^{n}}{M_{\Phi}^{n}}\partial hhhh\dots\dots\partial hh
\label{eq447}
\end{eqnarray}
From the amplitude we obtain 
$(G_{N})^{\frac{n}{2}} \sim\frac{1}{M_{\Phi}^{n}}$.

\section{I.T.B. Model Off-Shell}
\label{specul}
In the previous sections we have taken on-shell gravitons that satisfy the following relations :
\begin{eqnarray}
&&\Box h_{\mu\nu}=0 \nonumber \\
&&\partial^{\mu} h_{\mu\nu}=0 \nonumber \\
&&h^{\mu}\,_{\mu}=0 
 \label{eq450}
\end{eqnarray}
We can build the I.T.B model for off-shell gravitons too. In this case the rank $1, 3, 5$ fields couple to the following general currents : 
\begin{eqnarray}
&&J^{(1)}=a^{(1)}(\partial^{\mu}h_{\nu\rho})h^{\nu\rho}+b^{(1)}(\partial^{\nu}h_{\nu\rho})h^{\mu\rho}+c^{(1)}(\partial^{\mu}h^{\nu}_{\nu})h^{\rho}_{\rho}+d^{(1)}(\partial^{\nu}h_{\mu\rho})h_{\nu}^{\rho} \nonumber \\
&&J^{(3)}=a^{(3)}(\partial^{\mu}h^{\nu\sigma})h_{\sigma}^{\rho}+b^{(3)}(\partial^{\sigma}h^{\mu\nu})h_{\sigma}^{\rho}+c^{(3)}(\partial^{\sigma}h_{\sigma}^{\mu})h^{\nu\rho}+d^{(3)}(\partial^{\mu}h^{\sigma}_{\sigma})h^{\nu\rho}+e^{(3)}(\partial^{\mu}h^{\nu\rho})h^{\sigma}_{\sigma} \nonumber \\
&&J^{(5)}=a^{(5)}(\partial^{\mu}h^{\nu\rho})h^{\sigma\tau} \nonumber \\
\end{eqnarray}

\section{String Theory and ITB Model Toward a More \\Fundamental Theory}
In this section we analyse the possible connections between string theory, in the critical dimension $D=26$ for the Bosonic String  or $D=10$ for the Superstring, and the ITB Model introduced in the previous sections \cite{500}. In particular in the string theory contest it is simple to verify (at least in the bosonic case) that in the low energy limit $\alpha^{\prime} \rightarrow 0$ the effective vertexes for gravitons massive fields coupling contain only the tensor fields with rank $S=1, 3, 5$ as in the ITB Model.\\
The fondamental idea is the following:\\
at very high energy ($\alpha^{\prime} \rightarrow \infty$) we have a Gauge Theory of Massless Tensor Fields (G.T.H.S) that can lives in any dimension. In fact String Theory is Weyl anomaly free  in this regime as we can see from the Virasoro algebra (\ref{NoWeyl}). This fundamental theory could have many different Higgs' s phases related with many different relative minimum of complicated potential. The conjecture is that two of those are the String Theory and the I.T.B Model vacuums. In this picture we see that there is not a prefered dimension as in field theory, but the dimension is a consequence of the symmetry breaking. There is a vacuum where the massless fields of the microscopic theory become all massive with the exception of the gravitational multiplet and the spectrum is that of the String Theory so the dimension is D=10, and there is  another vacuum where the theory reproduces the gravitational theory in analogy with the Fermi theory of the weak interactions. The first vacuum is a finite theory (String Theory) and the other is a renormalizable gauge theory of Higher Spin in $D=4$ dimension with spontaneous symmetry breaking that contains only the graviton as massless state and a residual gauge symmetry that is the infinitesimal version of the $\mbox{Diff}$-invariance of General Relativity.  
\begin{equation}
[L_n , L_m] = (m-n) L_{n+m} + \frac{c}{12 \, \alpha^{\prime}} (m^3 - m) \delta_{m+n} \,\,\,\,\rightarrow\,\,\,\,[L_n , L_m] = (m-n) L_{n+m} 
\label{NoWeyl}
\end{equation}
Now we summarize the analogies between the Fermi Theory of weak interactions (F.T.W.I), General Relativity (G.R) and String Theory. At low energy the F.T.W.I model is a good approsimation of the weak interactions, it consists of a local vertex with four fermions and a fondamental constant $G_F$. At higher energies we see that the theory contains a vertex that consists of a fermionic current and a massive vector "W" (Intermediate Vector Boson Model, I.V.B). When we take the limit $k^2 \rightarrow 0$ we obtain the F.T.W.I with $G_F = 1/M_W$. In analogy we have introduced the Intermediate Tensor Boson Model (I.T.B) where now the fermionic current become a "pseudocurrent" with two gravitons and the W-bosons become fields with ramk $S=1, 3, 5$. Finally we have a theory very similar to string theory in the $\alpha^{\prime} \rightarrow 0$, in fact in this limit we obtain that the dominant Lagrangian terms obtained from scattering amplitudes in string theory are equal to the terms introduced in the I.T.B model.\\ 
It is likely there could be a microscopic theory of gravity similar to the standard model and we call this model :\\
$Gravity$ $Standard$ $Model$ (G.S.M), \\
where all the fields are massless and the big gauge invariance that includes all the fields produce a renormalizable theory in $4$-d.\\
On the other hand in the $\alpha^{\prime} \rightarrow \infty$ limit, String Theory can live in any dimension and it is likely that it is a theory for massless higher spin fields \cite{SF0}, \cite{SF1}, \cite{HSF}, \cite{M01}, \cite{VasilevEq}, \cite{200}, \cite{300}, \cite{400} that we call :\\
$Higher$ $Spin$ $Gauge$ $Theory$ (H.S.G.T).\\
In this pictures String Theory and the I.T.B theory are two different  phases of a more microscopic theory of Higher Spin Fields but it is possible there is another description of the microscopic Universe. \\
It is likely that String Theory and the I.T.B model are two different Hadronic phases of a Sub-Quark theory \cite{TRZ1}, \cite{TRZ2}, \cite{WETT}. At high energy we have an $SU(N)$ Yang-Mills Gauge Theory that contains $N_{f}$ Sub-Quarks in a free phase. On the other side at low energy in the confinement phase we obtain an Hadronic spectrum that reproduce all the string spectrum near a vacuum or the I.T.B model near another vacuum.

\section{Outlook and Conclusions}
In this paper we introduced a simple  model calling I.T.B (Intermediate Tensor Boson) for the description of interactions with four or more gravitons. We took Fermi theory of weak interactions as a starting point. In particular we focused on the evolution of the Fermi theory in the I.V.B theory (Intermediate Vector Boson). In our model the square root of $G_N$ constant is proportional to the inverse mediator mass $G_N \sim g/M_{\Phi}^{2}$, in analogy with the Fermi theory of weak interactions in which the square root of  Fermi constant is proportional to the inverse proton mass$G_F \sim g/M_{Proton}^{2}$.\\
In the four gravitons interaction case we introduced four Lagrangian terms that couple four "Gravitational Currents" with tensor fields of spin $1, 3, 5$. This is in analogy with the Fermi theory of weak interactions which couple the fermionic currents, containg two fermions,  with the Bosons $W^+$ and $W^-$. We can summary the analogy between the I.V.B model and I.T.B model in the following identifications
 \begin{eqnarray}
&h_{\mu\nu} \leftrightarrow \psi \nonumber \\
&\Phi_{\mu} \, , \, \Phi_{\mu\nu\rho} \, , \, \Phi_{\mu \nu \rho \sigma \tau} \leftrightarrow W^{+}_{\mu}\, , \, W^{-}_{\mu} \nonumber \\
&G_{N} \leftrightarrow G_{F}        
\end{eqnarray}
In order to obtain the local interactions of  Einstein action we introduced tensor field propagators  that are reducible representations of the Poincar\'e group (see section \ref{RedPoin}).\\   
We write the complete interaction Lagrangian for four gravitons interactions 
\begin{eqnarray}
L_I = g(\partial^{\mu}h_{\nu \rho}) h^{\nu \rho} \Phi_{\mu}\nonumber + 
g(\partial^{\mu}h^{\nu\sigma})h_{\sigma}\,^{\rho}\Phi_{\mu\nu\rho} \nonumber + g(\partial^{\sigma}h^{\mu\nu})h_{\sigma}\,^{\rho}\Phi_{\mu\nu\rho}\nonumber +
g(\partial^{\mu}h^{\nu\rho})h^{\sigma\tau}\Phi_{\mu\nu\rho\sigma\tau}
\end{eqnarray}
 We also studied the interactions with $5, 6, \dots n$ gravitons and we showed that the tensor structure can be reproduced in the I.T.B model using another Lagrangian term that contain a  
tensor field and two graviton without derivative.  We write the complete interaction Lagrangian  reproducing the tensor structure of all the interactions with $5, 6, \dots n$ gravitons
 \begin{eqnarray}
L_I & = & g(\partial^{\mu}h_{\nu \rho}) h^{\nu \rho} \Phi_{\mu}\nonumber +
g(\partial^{\mu}h^{\nu\sigma})h_{\sigma}\,^{\rho}\Phi_{\mu\nu\rho} \nonumber + g(\partial^{\sigma}h^{\mu\nu})h_{\sigma}\,^{\rho}\Phi_{\mu\nu\rho}\nonumber +
g(\partial^{\mu}h^{\nu\rho})h^{\sigma\tau}\Phi_{\mu\nu\rho\sigma\tau} \nonumber \\
& + &g^{\prime}M_{\Phi}h_{\mu\nu}\Phi^{\mu}\Phi^{\nu} +
gM_{\Phi}h_{\mu\nu}\Phi^{\mu\rho\sigma}\Phi^{\nu}\, _{\rho\sigma}+ g^{\prime \prime} M_{\Phi}h_{\mu\nu}\Phi^{\mu\rho\sigma\tau\delta}\Phi^{\nu}\, _{\rho\sigma\tau\delta} 
\end{eqnarray}
In this simple model the coupling constants are all dimensionless constants and the Newton constant is connected with the mediators mass $G_N \sim 1/M_{\Phi}^{2}$. Although the theory is not renormalizable because the mediators are higher spin fields. \\
 As we said in the section \ref{specul} it is likely there is a theory of gravity similar to the standard model for weak and electro-magnetic interactions. This more fundamental theory could be a gauge theory of higher spin fields and so the Quantum Gravity would be a possible broken phase of this theory. Another possible phase of our new theory could be a string theory as we said in \ref{specul}.  \\
 Another possibility is that there is an alternative more fundamental theory (\ref{specul}). It is likely that at high energy level we have an $SU(N)$ Yang-Mills Gauge Theory that contain $N_{f}$ Sub-Quark in a free phase. On the other side at low energy in the confinement phase we obtain an hadronic spectrum that reproduce the I.T.B model near a vacuum or all the string spectrum near another vacuum.

\section*{Acknowledgements}
We are grateful to Enore Guadagnini, Eugenio Bianchi, Gabriele Marchi and Giuseppe Tarabella for many important and clarifying discussions.

\newpage
\appendix
\section{Explicit Calculation for  $4$ Gravitons Interaction}
\markright{CALCOLO DETTAGLIATO DELL'AMPIEZZA...}
\begin{eqnarray}
&&S_{(2)} =\frac{(i)^{2}}{2!} \int d^{4}x_{1}d^{4}x_{2} T[:(L_{(1)}+L_{(2)}+L_{(3)}+L_{(4)})(x_{1}): \nonumber \\
      &&\hspace{3.7cm}:(L_{(1)}+L_{(2)}+L_{(3)}+L_{(4)})(x_{2}):]= \nonumber \\
&&=\frac{(i)^{2}}{2!} \int d^{4}x_{1}d^{4}x_{2} T[:(L_{(1)}+L_{(2)}):(x_{1})::(L_{(1)}+L_{(2)}):(x_{2}):]+ \nonumber \\
&&\hspace{2.9cm}+T[:L_{(3)}(x_{1})::L_{(3)}(x_{2}):]+T[:L_{(4)}(x_{1})::L_{(4)}(x_{2}):]+ \nonumber \\
&&\hspace{2.9cm}+gravitons-contraction= \nonumber \\
&&=\frac{(i)^{2}}{2!} \int d^{4}x_{1}d^{4}x_{2}[g^{2}:(J_{(1)}^{\mu_{1}\nu_{1}\rho_{1}} \WT{\Phi_{\mu_{1}\nu_{1}\rho_{1}})(x_{1}):: (J_{(1)}^{\mu_{2}\nu_{2}\rho_{2}}\Phi }\hspace{-0.1cm}\vphantom{.}_{\mu_{2}\nu_{2}\rho_{2}})(x_{2}):+ \nonumber \\
&&\hspace{2.8cm}+2g^{2}:(J_{(1)}^{\mu_{1}\nu_{1}\rho_{1}} \WT{\Phi_{\mu_{1}\nu_{1}\rho_{1}})(x_{1}):: (J_{(2)}^{\mu_{2}\nu_{2}\rho_{2}}\Phi }\hspace{-0.1cm}\vphantom{.}_{\mu_{2}\nu_{2}\rho_{2}})(x_{2}):+ \nonumber \\
&&\hspace{3.0cm}+g^{2}:(J_{(2)}^{\mu_{1}\nu_{1}\rho_{1}} \WT{\Phi_{\mu_{1}\nu_{1}\rho_{1}})(x_{1}):: (J_{(2)}^{\mu_{2}\nu_{2}\rho_{2}}\Phi }\hspace{-0.1cm}\vphantom{.}_{\mu_{2}\nu_{2}\rho_{2}})(x_{2}):+ \nonumber \\
&&\hspace{0.6cm}+g^{2}:(J_{(3)}^{\mu_{1}\nu_{1}\rho_{1}\sigma_{1}\tau_{1}} \WT{\Phi_{\mu_{1}\nu_{1}\rho_{1}\sigma_{1}\tau_{1}})(x_{1}):: (J_{(3)}^{\mu_{2}\nu_{2}\rho_{2}\sigma_{2}\tau_{2}}\Phi }\hspace{-0.1cm}\vphantom{.}_{\mu_{2}\nu_{2}\rho_{2}\sigma_{2}\tau_{2}})(x_{2}):+ \nonumber \\
&&\hspace{5.4cm}+g^{2}:(J_{(4)}^{\mu_{1}} \WT{\Phi_{\mu_{1}})(x_{1}):(J_{(4)}^{\mu_{2}}\Phi}\hspace{-0.1cm}\vphantom{.}_{\mu_{2}})(x_{1}):+ \nonumber \\
&&\hspace{6.8cm}+ gravitons-contraction= \nonumber \\
&&=\frac{(i)^{2}}{2!} \int d^{4}x_{1}d^{4}x_{2}[g^{2}J_{(1)}^{\mu_{1}\nu_{1}\rho_{1}} (x_{1}) \int \frac{d^{4}k_{1}}{(2\pi)^{4}}(-i)\frac{e^{ik_{1}(x_{2}-x_{1})}}{k_{1}^{2}-M_{\Phi}^{2}}K^{(3)}_{\mu_{1}\nu_{1}\rho_{1}\, , \,\mu_{2}\nu_{2}\rho_{2}} \times \nonumber \\
&&\hspace{3.0cm}\times J_{(1)}^{\mu_{2}\nu_{2}\rho_{2}}(x_{2})+ \nonumber \\
&&\hspace{3.0cm}+2g^{2}J_{(1)}^{\mu_{1}\nu_{1}\rho_{1}} (x_{1}) \int \frac{d^{4}k_{1}}{(2\pi)^{4}}(-i)\frac{e^{ik_{1}(x_{2}-x_{1})}}{k_{1}^{2}-M_{\Phi}^{2}}K^{(3)}_{\mu_{1}\nu_{1}\rho_{1}\, , \,\mu_{2}\nu_{2}\rho_{2}} \times \nonumber  \\
&&\hspace{3.0cm}\times J_{(2)}^{\mu_{2}\nu_{2}\rho_{2}}(x_{2})+ \nonumber \\
&&\hspace{3.0cm}+g^{2}J_{(2)}^{\mu_{1}\nu_{1}\rho_{1}} (x_{1}) \int \frac{d^{4}k_{1}}{(2\pi)^{4}}(-i)\frac{e^{ik_{1}(x_{2}-x_{1})}}{k_{1}^{2}-M_{\Phi}^{2}}K^{(3)}_{\mu_{1}\nu_{1}\rho_{1}\, , \,\mu_{2}\nu_{2}\rho_{2}} \times \nonumber \\
&&\hspace{3.0cm}\times J_{(2)}^{\mu_{2}\nu_{2}\rho_{2}}(x_{2})+ \nonumber \\
&&\hspace{3.0cm}+g^{2}J_{(3)}^{\mu_{1}\nu_{1}\rho_{1}\sigma_{1}\tau_{1}} (x_{1}) \int \frac{d^{4}k_{2}}{(2\pi)^{4}}(-i)\frac{e^{ik_{2}(x_{2}-x_{1})}}{k_{1}^{2}-M_{\Phi}^{2}}\times \nonumber \\
&&\hspace{3.0cm}\times K^{(5)}_{\mu_{1}\nu_{1}\rho_{1}\sigma_{1}\tau_{1}\, , \,\mu_{2}\nu_{2}\rho_{2}\sigma_{2}\tau_{2}} J_{(3)}^{\mu_{2}\nu_{2}\rho_{2}\sigma_{2}\tau_{2}}(x_{2})+ \nonumber \\
&&\hspace{3.0cm}+g^{2}J_{(4)}^{\mu_{1}}(x_{1})\int \frac{d^{4}k_{3}}{(2\pi)^{4}}(-i)\frac{e^{ik_{3}(x_{2}-x_{1})}}{k_{3}^{2}-M_{\Phi}^{2}}K^{(1)}_{\mu_{1}\, , \,\mu_{2}}J_{(4)}^{\mu_{2}}(x_{2})\rightarrow \nonumber \\
\label{eq443}
\end{eqnarray}
We continue the calculation in the limit $M_{\Phi}\rightarrow\infty$, using the tensor \ref{eq437}, \ref{eq438}, \ref{eq439} :
\begin{eqnarray}
&&\rightarrow \frac{(-i)}{2M_{\Phi}^{2}}g^{2} \int d^{4}x_{1}[\frac{1}{2} (\partial^{\mu_{1}}h^{\nu_{1}\sigma_{1}})h_{\sigma_{1}}\,^{\rho_{1}}(\partial_{\mu_{1}}h_{\nu_{1}}\,^{\sigma_{2}})h_{\sigma_{2}\rho_{1}}+ \nonumber \\
&&\hspace{2.5cm}+\frac{3}{2} (\partial^{\mu_{1}}h^{\nu_{1}\sigma_{1}})h_{\sigma_{1}}\,^{\rho_{1}}(\partial_{\mu_{1}}h_{\rho_{1}}\,^{\sigma_{2}})h_{\sigma_{2}\nu_{1}}+ \nonumber \\
&&\hspace{2.5cm}+ (\partial^{\mu_{1}}h^{\nu_{1}\sigma_{1}})h_{\sigma_{1}}\,^{\rho_{1}}(\partial_{\nu_{1}}h_{\mu_{1}}\,^{\sigma_{2}})h_{\sigma_{2}\rho_{1}}+ \nonumber \\
&&\hspace{2.5cm}-\frac{15}{8} (\partial^{\mu_{1}}h^{\nu_{1}\sigma_{1}})h_{\sigma_{1}}\,^{\rho_{1}}(\partial_{\nu_{1}}h_{\rho_{1}}\,^{\sigma_{2}})h_{\sigma_{2}\mu_{1}}+ \nonumber \\
&&\hspace{2.5cm}+\frac{5}{8} (\partial^{\mu_{1}}h^{\mu_{1}\sigma_{1}})h_{\sigma_{1}}\,^{\rho_{1}}(\partial_{\rho_{1}}h_{\mu_{1}}\,^{\sigma_{2}})h_{\sigma_{2}\nu_{1}}+ \nonumber \\
&&\hspace{2.5cm}+\frac{7}{8} (\partial^{\mu_{1}}h^{\nu_{1}\sigma_{1}})h_{\sigma_{1}}\,^{\rho_{1}}(\partial_{\rho_{1}}h_{\nu_{1}}\,^{\sigma_{2}})h_{\sigma_{2}\mu_{1}}+ \nonumber \\
&&\hspace{2.5cm}+ (\partial^{\mu_{1}}h^{\nu_{1}\sigma_{1}})h_{\sigma_{1}}\,^{\rho_{1}}(\partial^{\sigma_{2}}h_{\mu_{1}\nu_{1}})h_{\sigma_{2}\rho_{1}}+ \nonumber \\
&&\hspace{2.5cm}+3 (\partial^{\mu_{1}}h^{\nu_{1}\sigma_{1}})h_{\sigma_{1}}\,^{\rho_{1}}(\partial^{\sigma_{2}}h_{\mu_{1}\rho_{1}})h_{\sigma_{2}\nu_{1}}+ \nonumber \\
&&\hspace{2.5cm}+2 (\partial^{\mu_{1}}h^{\nu_{1}\sigma_{1}})h_{\sigma_{1}}\,^{\rho_{1}}(\partial^{\sigma_{2}}h_{\nu_{1}\mu_{1}})h_{\sigma_{2}\rho_{1}}+\nonumber \\
&&\hspace{2.5cm}-\frac{15}{4} (\partial^{\mu_{1}}h^{\nu_{1}\sigma_{1}})h_{\sigma_{1}}\,^{\rho_{1}}(\partial^{\sigma_{2}}h_{\nu_{1}\rho_{1}})h_{\sigma_{2}\mu_{1}}+\nonumber \\
&&\hspace{2.5cm}+\frac{5}{4} (\partial^{\mu_{1}}h^{\nu_{1}\sigma_{1}})h_{\sigma_{1}}\,^{\rho_{1}}(\partial^{\sigma_{2}}h_{\rho_{1}\mu_{1}})h_{\sigma_{2}\nu_{1}}+ \nonumber \\
&&\hspace{2.5cm}+\frac{7}{4} (\partial^{\mu_{1}}h^{\nu_{1}\sigma_{1}})h_{\sigma_{1}}\,^{\rho_{1}}(\partial^{\sigma_{2}}h_{\rho_{1}\nu_{1}})h_{\sigma_{2}\mu_{1}}+ \nonumber \\
&&\hspace{2.5cm}+\frac{1}{2} (\partial^{\sigma_{1}}h^{\mu_{1}\nu_{1}})h_{\sigma_{1}}\,^{\rho_{1}}(\partial^{\sigma_{2}}h_{\mu_{1}\nu_{1}})h_{\sigma_{2}\rho_{1}}+\nonumber \\
&&\hspace{2.5cm}+\frac{3}{2} (\partial^{\sigma_{1}}h^{\mu_{1}\nu_{1}})h_{\sigma_{1}}\,^{\rho_{1}}(\partial^{\sigma_{2}}h_{\mu_{1}\rho_{1}})h_{\sigma_{2}\nu_{1}}+ \nonumber \\
&&\hspace{2.5cm}+(\partial^{\sigma_{1}}h^{\mu_{1}\nu_{1}})h_{\sigma_{1}}\,^{\rho_{1}}(\partial^{\sigma_{2}}h_{\nu_{1}\mu_{1}})h_{\sigma_{2}\rho_{1}}+ \nonumber \\
&&\hspace{2.5cm}+\frac{5}{8} (\partial^{\sigma_{1}}h^{\mu_{1}\nu_{1}})h_{\sigma_{1}}\,^{\rho_{1}}(\partial^{\sigma_{2}}h_{\nu_{1}\rho_{1}})h_{\sigma_{2}\mu_{1}}+ \nonumber \\
&&\hspace{2.5cm}-\frac{15}{8} (\partial^{\sigma_{1}}h^{\mu_{1}\nu_{1}})h_{\sigma_{1}}\,^{\rho_{1}}(\partial^{\sigma_{2}}h_{\rho_{1}\mu_{1}})h_{\sigma_{2}\nu_{1}}+\nonumber \\
&&\hspace{2.5cm}+\frac{7}{8} (\partial^{\sigma_{1}}h^{\mu_{1}\nu_{1}})h_{\sigma_{1}}\,^{\rho_{1}}(\partial^{\sigma_{2}}h_{\rho_{1}\nu_{1}})h_{\sigma_{2}\mu_{1}}+\nonumber \\
&&\hspace{2.5cm}+\frac{5}{2}h^{\mu\nu}\partial^{\tau}h^{\rho\sigma}h_{\mu\sigma}\partial_{\tau}h_{\rho\nu}-\frac{5}{2}h^{\mu\nu}\partial^{\tau}h^{\rho\sigma}h_{\mu\tau}\partial_{\nu}h_{\rho\tau}+ \nonumber \\
&&\hspace{2.5cm}+\frac{13}{2}h^{\mu\nu}\partial^{\tau}h^{\rho\sigma}h_{\mu\nu}\partial_{\tau}h_{\rho\sigma}+\frac{1}{2}h^{\mu\nu}\partial^{\tau}h^{\rho\sigma}h_{\mu\nu}\partial_{\sigma}h_{\rho\tau}+ \nonumber \\
&&\hspace{2.5cm}+\frac{29}{2}h^{\mu\nu} \partial_{\tau} h_{\mu\nu} h^{\rho\sigma}\partial^{\tau}  h_{\rho\sigma}]
\label{eq444}
\end{eqnarray}
The first $18$ lines concern the rank $3$ field, the lines $19$ and $20$ concern the rank $5$ and the final line concerns the spin $1$ field. Now we rewrite the result summing the equal terms in the first $18$ lines.
\begin{eqnarray}
\frac{(-i)}{2M_{\Phi}^{2}}g^{2} \int d^{4}x \hspace{-0.7cm} &&[2h^{\mu\nu}\partial_{\tau}h_{\nu\rho}h^{\tau\sigma}\partial_{\mu}h_{\sigma}\,^{\rho}-2h^{\mu\nu}\partial_{\tau}h_{\nu\rho}h^{\tau\sigma}\partial_{\sigma}h_{\mu}\,^{\rho}+\nonumber \\
&&+3h^{\mu\nu}\partial_{\tau}h_{\nu\rho}h^{\rho\sigma}\partial_{\mu}h^{\tau}\,_{\sigma}+\frac{3}{2}h^{\mu\nu}\partial_{\tau}h_{\nu\rho}h^{\rho\sigma}\partial^{\tau}h_{\mu\sigma}+ \nonumber \\
&&+2h^{\mu\nu}\partial_{\tau}h_{\nu\rho}h_{\mu\sigma}\partial^{\sigma}h^{\tau\rho}+h^{\mu\nu}\partial_{\tau}h_{\mu\rho}h_{\nu\sigma}\partial^{\rho}h^{\tau\sigma}+ \nonumber \\
&&+\frac{1}{2}h^{\mu\nu}\partial_{\tau}h_{\mu\rho}h_{\nu\sigma}\partial^{\tau}h^{\rho\sigma}+h^{\mu\nu}\partial_{\mu}h_{\tau\rho}h_{\nu\sigma}\partial^{\rho}h^{\tau\sigma}+\nonumber \\
&&+\frac{3}{2}h^{\mu\nu}\partial_{\mu}h_{\tau\rho}h_{\nu\sigma}\partial^{\sigma}h^{\tau\rho}+\nonumber \\
&&+\frac{5}{2}h^{\mu\nu}\partial^{\tau}h^{\rho\sigma}h_{\mu\sigma}\partial_{\tau}h_{\rho\nu}-\frac{5}{2}h^{\mu\nu}\partial^{\tau}h^{\rho\sigma}h_{\mu\tau}\partial_{\nu}h_{\rho\tau}+\nonumber \\
&&+\frac{13}{2}h^{\mu\nu}\partial^{\tau}h^{\rho\sigma}h_{\mu\nu}\partial_{\tau}h_{\rho\sigma}+\frac{1}{2}h^{\mu\nu}\partial^{\tau}h^{\rho\sigma}h_{\mu\nu}\partial_{\sigma}h_{\rho\tau}+\nonumber \\
&&+\frac{29}{2}h^{\mu\nu} \partial_{\tau} h_{\mu\nu} h^{\rho\sigma}\partial^{\tau}  h_{\rho\sigma}]
\label{eq445}
\end{eqnarray}
If we want obtain the forth order term of General Relativity from the I.T.B model we must have that 
$\frac{g^{2}}{2M_{\Phi}^{2}}=64\pi G_{N}$ and so 
$M_{\Phi}^{2}=\frac{g^{2}}{128\pi G_{N}}$.

\end{document}